\RequirePackage{fix-cm}
\documentclass[twocolumn,epjc3]{svjour3}  

\smartqed  

\usepackage{graphicx}  
\usepackage{dcolumn}   
\usepackage{bm}        
\usepackage{amssymb}   
\usepackage{sidecap}   
\usepackage{subfigure}
\usepackage{amsmath}
\usepackage{multirow}
\usepackage{xspace}
\usepackage{booktabs}
\usepackage[normalem]{ulem}
\usepackage[numbers,sort&compress]{natbib}
\usepackage{epstopdf} 

\usepackage{hyperref}

\usepackage{atlasphysics} 
\usepackage{subfigure}
\usepackage{multirow}

\usepackage{preprintcover}  
\PreprintCoverPaperTitle{Search for a light charged Higgs boson in the decay channel $H^+ \rightarrow c\bar{s}$ in $\ttbar$ events using $pp$ collisions at $\sqrt{s} = 7$~TeV with the ATLAS detector} 
\PreprintIdNumber{CERN-PH-EP-2012-338}  
\PreprintCoverAbstract{A search for a charged Higgs boson ($H^+$) in $\ttbar$ decays is presented, where one of the top 
quarks decays via $t\rightarrow H^+b$, followed by $H^+\rightarrow$ two jets ($\csbar$). The other
top quark decays to $Wb$, where the $W$~boson then decays into a lepton ($e/\mu$) and a neutrino.
The data were recorded in $pp$ collisions at $\sqrt{s} = 7$~TeV by the ATLAS detector 
at the LHC in 2011, and correspond to an integrated luminosity of 4.7 fb$^{-1}$.
With no observation of a signal, 
 95\% confidence level (CL) upper limits are set on the decay branching ratio 
 of top quarks to charged Higgs bosons varying between 5\% and 1\%
for $H^+$ masses between 90 GeV and 150 GeV, assuming ${\cal B}(H^+ \rightarrow c\bar{s})=100\%$.
}  
\PreprintJournalName{Eur. Phys. J. C}  

\def\met{\ensuremath{E_{\mathrm{T}}^{\mathrm{miss}}}}
\def \csbar    {c\bar{s}}

\def \ttbar    {t\bar{t}}
\def \bbbar    {b\bar{b}}

\def\mt2{\ensuremath{m_{T2}^{H^{\pm}}}}
\def\mtp2{\ensuremath{m_{T2}^{H^{+}}}}
\def\Br{\ensuremath{\mathcal B}}
\def\pT{\ensuremath{p_{\mathrm{T}}}}

\journalname{Eur.\ Phys.\ J.\ C}

\begin{document}

\title{Search for a light charged Higgs boson  
in the decay channel $H^+ \rightarrow c\bar{s}$ in $\ttbar$ events  using $pp$ collisions at $\sqrt{s} = 7$~TeV with the ATLAS detector}

\author{The ATLAS Collaboration}
\institute{CERN, 1211 Geneva 23, Switzerland}
\date{February 14, 2013}

\maketitle 

\begin{abstract}
A search for a charged Higgs boson ($H^+$) in $\ttbar$ decays is presented, where one of the top 
quarks decays via $t\rightarrow H^+b$, followed by $H^+\rightarrow$ two jets ($\csbar$). The other
top quark decays to $Wb$, where the $W$~boson then decays into a lepton ($e/\mu$) and a neutrino.
The data were recorded in $pp$ collisions at $\sqrt{s} = 7$~TeV by the ATLAS detector 
at the LHC in 2011, and correspond to an integrated luminosity of 4.7 fb$^{-1}$.
With no observation of a signal, 
 95\% confidence level (CL) upper limits are set on the decay branching ratio 
 of top quarks to charged Higgs bosons varying between 5\% and 1\%
for $H^+$ masses between 90 GeV and 150 GeV, assuming ${\cal B}(H^+ \rightarrow c\bar{s})=100\%$.
\smallskip
\flushleft{\bf PACS} 12.60.Fr,  14.65.Ha, 14.80.Da, 14.80.Fd
\end{abstract}
\section{Introduction}

In the Standard Model (SM), electroweak symmetry breaking (EWSB)
occurs through a single complex 
scalar doublet field and results in a single physical state, the Higgs boson~\cite{EnglertBrout,Higgs,Guralnik}. A particle with characteristics of the SM Higgs boson has been discovered by both ATLAS~\cite{HiggsATLAS} and CMS~\cite{HiggsCMS}.
Beyond the SM, many models have been proposed, extending the Higgs sector to explain EWSB.
The newly discovered boson is compatible with many of these models so that discovering its true nature is crucial to understanding EWSB.
Two Higgs-doublet models (2HDM)~\cite{2hdm} are simple extensions of the SM with five observable Higgs bosons, of which two are charged ($H^{+}$ and $H^{-}$) and three are neutral~($h^0$, $H^0$ and $A^0$). The discovery of a charged Higgs 
boson would be a signal for new physics beyond the SM.

The Minimal Supersymmetric Standard Model\\(MSSM)~\cite{MSSM} is an example of a 2HDM. 
At tree level, the MSSM Higgs sector is determined by two independent 
parameters, which can be taken to be the mass 
$m_{H^+}$
 and the ratio of the two Higgs doublet vacuum expectation values,
parametrized by $\tan\beta$. In the MSSM, 
a light $H^+$ (defined as $m_{H^+}<m_t$) decays predominantly to $c\bar{s}$, $b\bar{b}W^+$, and 
$\tau^+ \nu$, with the respective branching ratios depending on $\tan \beta$ and $m_{H^+}$. Charge conjugated processes are implied throughout this paper.
For $\tan\beta <1$, $c\bar{s}$~is an important decay mode with 
${\cal B}(H^+ \rightarrow c\bar{s})$ near 70\%~\cite{Heinemeyer:1998yj,Lee2009} for $m_{H^\pm} \simeq 110$~GeV, whereas 
for $\tan \beta>3$, $H^+ \rightarrow \tau^+ \nu$ dominates (90\%). For higher $H^+$ masses at low $\tan\beta$, the decay mode $H^+\rightarrow Wb\bar{b}$ can be dominant.
A light MSSM charged Higgs boson is viable at a relatively low $\tan\beta \approx 6$
in certain MSSM benchmark scenarios~\cite{Carena:2013qia} that take into account the discovery
of a Higgs boson with a mass of 125~GeV at the LHC.

The LEP experiments placed lower limits on $m_{H^+}$ in any type-II 2HDM~\cite{Type2} 
varying between 75 GeV and 91~GeV~\cite{mssmlep1,mssmlep2,mssmlep3,Abbiendi:2008aa,mssmlep5} depending on the assumed decay branching 
ratios for the charged Higgs boson.
At the Tevatron, searches for charged Higgs bosons have been
extended to larger values of $m_{H^+}$.
No evidence for a $H^+$ was found and upper limits were set on 
the branching ratio $\Br(t \rightarrow H^{+}b)$ varying between $10\%$ and $30\%$
for a light $H^+$ under the assumption of $\Br(H^+ \rightarrow c\bar{s})=100\%$~\cite{cdfRef,dzeroRef}.
The discovery of a Higgs boson at the LHC is a weak constraint on many 2HDMs, 
and is compatible with the existence of a light charged Higgs decaying to two jets, 
especially in type I 2HDMs \cite{PhysRevD.85.077703,Craig:2012vn}.

In this paper, a search for a charged Higgs boson produced
in $\ttbar$ decays is presented, where
one of the top quarks decays via $t \rightarrow H^+b$ 
with the charged Higgs boson subsequently decaying to two jets ($\csbar$), where again a $100\%$ branching fraction is assumed.
The other top quark decays according to the SM via $\bar{t} \rightarrow W^-\bar{b}$~with
the $W$~boson decaying into a lepton ($e/\mu$) and the corresponding neutrino.
The signal process therefore has the same topology as SM $\ttbar$ decays in the lepton + jets channel, where one $W$ decays to two jets and the other to a lepton and corresponding neutrino, but the invariant 
mass of the two jets from the $H^+$ peaks at $m_{H^{+}}$.
The search is performed by comparing the dijet mass spectrum in the data
with the prediction from SM top-quark decays and with the expectation 
of a top quark having a non-zero branching ratio for decay to $H^+b$.

\section{Detector description and event samples}

The data used in the analysis were recorded by the ATLAS detector in proton--proton ($pp$) collisions at a centre-of-mass energy of $\sqrt{s}=7$~TeV during the 2011 data-taking period of the Large Hadron Collider (LHC) \cite{LHC}. Events were required to pass a high-transverse momentum ($\pt$) single-lepton ($e/\mu$) trigger, and to have been recorded when
all detector systems critical to muon, electron, and jet identification were operational.
The lepton triggers required in the different data taking periods had varying $\pt$ thresholds:  20--22~GeV for the electron trigger and 18~GeV for the muon trigger.
The resulting dataset corresponds to an integrated luminosity of 4.7~fb$^{-1}$~\cite{Aad:2011dr,ATLAS-CONF-2011-116}. 

The ATLAS detector~\cite{Aad:2008zzm} consists of an inner tracking system immersed in a $2$ T axial magnetic
field provided by a thin solenoid; electromagnetic and hadronic calorimeters;
and a muon spectrometer (MS) embedded in a toroidal magnet system.
The inner detector tracking system (ID) comprises a silicon pixel detector closest to the beamline, a silicon microstrip detector, and a straw tube transition radiation tracker.
The electromagnetic (EM) calorimeters are high-granularity liquid-argon sampling calorimeters with lead as the absorber material in the barrel and endcap regions, and copper in the forward region.
The hadronic \hyphenation{calor-imetry}calorimetry uses two different detector technologies. The barrel calorimeter $\left(\left|\eta\right|<1.7\right)$\footnote{ATLAS uses a right-handed coordinate system with its origin at the nominal interaction point (IP) in the centre of the detector and the $z$-axis along the beam pipe. The $x$-axis points from the IP to the centre of the LHC ring, and the $y$-axis points upward. Cylindrical coordinates $(r,\phi)$ are used in the transverse ($x$, $y$) plane, $\phi$ being the azimuthal angle around the beam pipe. The pseudorapidity is defined in terms of the polar angle $\theta$ as $\eta=-\ln\tan(\theta/2)$.} consists of scintillator tiles interleaved with steel absorber plates. The endcap $(1.5 < | \eta| <3.2)$ and forward $(3.1< |\eta|<4.9)$ calorimeters both use liquid argon as the active material, and copper and tungsten respectively as the absorber.
The MS consists of three large superconducting toroids each with eight coils,
and a system of precision tracking and fast trigger chambers.

The largest background to the charged Higgs boson signal is the SM production and decay of $\ttbar$ pairs. 
Additional background contributions (referred to as non-$\ttbar$ backgrounds)
arise from the production of a single top quark, of a $W$ or $Z$ boson with additional jets, of QCD multi-jets, and of dibosons.

Top-quark pair and single top-quark events ($Wt$-channel and $s$-channel) were generated using the {\sc mc@n-lo 4.01} \cite{mcatnlo1,mcatnlo2,Frixione:2005vw,Frixione:2008yi} Monte Carlo (MC) generator 
coupled to {\sc Herwig~6.520.2}~\cite{herwig} to provide the parton showering
and hadronization using the {\sc AUET2-CT10} \cite{HerwigJimmyTune2011,Lai:2010vv} tune; {\sc Jimmy} \cite{jimmy} was used to model the underlying event. Single top-quark events in the $t$-channel were generated using {\sc AcerMC~3.8} \cite{Kersevan:2004yg} coupled to {\sc Pythia 6.425} \cite{Pythia} with the {\sc AUET2-MRST2007LO**} \cite{HerwigJimmyTune2011,Sherstnev:2007nd} tune.
$W/Z$+jet and diboson events were generated using the leading-order (LO) {\sc Alpgen 2.13}~\cite{Mangano:2002e} generator interfaced to {\sc Herwig} with the {\sc AUET2-CTEQ6L1} \cite{HerwigJimmyTune2011,Nadolsky:2008zw} tune. The $W/Z$+jet simulated data include dedicated samples for heavy-flavour production ($b\bar{b}, c\bar{c}$ and $c$).
Signal samples of $\ttbar\to H^+bW^-\bar{b}$ were generated 
using {\sc Pythia~6.425}
for seven different $H^+$ masses from 90 GeV to 150 GeV.

The data are affected by the detector response to multiple $pp$ interactions occurring in the same or neighbouring bunch crossings, known as pile-up. Minimum-bias interactions generated by {\sc åÊPythia 6.425} \cite{Pythia}, which has been tuned to data \cite{Aad:2010ac}, were overlaid on the simulated signal and background events.
The events were weighted to reproduce the distribution of the number of interactions per bunch crossing observed in the data.
A {\sc Geant4} simulation \cite{Agostinelli:2002hh,:2010wqa} is used to model the response of the ATLAS detector, and the samples are reconstructed and analysed in the same way as the data.

\section{Physics objects and event selection}

Jets are reconstructed from topological clusters of calori-meter cells~\cite{Topo} using the anti-$k_t$~algorithm~\cite{Cacciari:2008gp,Cacciari200657} with
a radius parameter $R=0.4$. Topological clusters are built using an algorithm that suppresses detector noise.
Jets are corrected back to particle (truth) level using calibrations derived from Monte Carlo simulation and validated
with both test-beam~\cite{Abat:2010zz} and collision-data studies~\cite{Aad:2011he}.
Events are excluded if they contain a high-$\pt$ jet that fails quality criteria
rejecting detector noise and non-collision backgrounds~\cite{atlasJetQual}.
To suppress the use of jets originating from secondary $pp$ interactions, a jet vertex fraction (JVF) algorithm is used. Inner detector tracks, with $\pt >1$~GeV, are uniquely associated with jets using $\Delta R(\mathrm{jet, track}) < 0.4$, where $\Delta R\equiv\sqrt{( \Delta\phi )^2 + ( \Delta\eta )^2}$. The JVF algorithm requires that at least $75\%$ of the sum of the $\pt$ of the tracks associated with the jet is from tracks compatible with originating from the primary vertex of the event.
Tagging algorithms identify jets originating from $b$-quark decays by selecting jets with tracks from
secondary vertices or those with a large impact parameter significance. A multivariate algorithm (MV1)~\cite{MV1-12}, which
uses a neural network to combine the weights from multiple tagging algorithms, is used to identify jets originating from $b$-quarks. 
Jets passing the MV1 selection are referred to as $b$-tagged jets.
The selection on the discriminating variable of the algorithm achieves an average per-jet efficiency of $70\%$~to select $b$-jets in $\ttbar$ events, with a probability to incorrectly tag light jets of less than $0.1\%$~\cite{MV1-2}. Studies have shown that this working point has a 20--40$\%$ efficiency to tag a $c$-jet, depending on the $\pt$ of the jet~\cite{MV1-123}.

Muons are required to be identified in both the ID and MS, and their momentum is obtained through a combined
fit of all hits in both systems.
Muons are also required to satisfy isolation criteria to reject those originating from heavy-flavour decays 
and hadrons misidentified as muons. The sum of the transverse momenta of ID tracks within a cone of
$\Delta R = 0.3$~around the muon, excluding the muon track itself, is required to be less
than $2.5$~GeV.
The transverse energy measured in the calorimeters within a cone of $\Delta R=0.2$,~excluding the energy
associated with the muon, is required to be less than $4$~GeV. In addition, muons are removed if they are found within $\Delta R<0.4$ of a jet that has $\pt > 25$~GeV~\cite{TopXSec,ATLAS-CONF-2010-036}.

The reconstruction of electron candidates starts from a seed cluster in the second layer
of the EM calorimeter. The cluster is matched to a track found in the ID and a set of selection criteria 
are applied to reject electron candidates originating from jets~\cite{Aad:2010yt}. Electrons are required to be isolated
in order to suppress the QCD multi-jet background.
The calorimeter isolation is performed using a cone of $\Delta R=0.2$ and
the track isolation uses a cone of radius $\Delta R=0.3$.
The calorimeter and track isolation cut values are chosen to achieve 
90\% efficiency with respect to selected electron candidates~\cite{Aad:2011mk}.
As in the case of muons, the electron itself is excluded from the sum over the isolation cone.

Energy deposits in the calorimeter are expressed 
as four-vectors $(E,\vec{p})$, 
where the direction is determined from the position of the calorimeter cluster
 and the nominal interaction point ($x=y=z=0$). The clusters are formed assuming $E=|p|$.
The missing transverse momentum ($\met$) is given 
by the negative of the vector sum of the calorimeter four-momenta, projected 
into the $(x, y)$ plane. The $\met$ calculation uses the energy scale 
appropriate for each physics object described above. 
For muons, the momentum measured from the combined tracking
is used as the energy.
The remaining calorimeter cells not associated 
with any physics object are included 
at the electromagnetic energy scale of the calorimeter~\cite{Aad:2012re}.

A set of requirements is imposed to select events containing $\ttbar$ decays in the lepton$+$jets channel~\cite{TopXSec}.
First, events are required to contain a primary vertex with at least five associated tracks to suppress non-collision backgrounds.
Exactly one electron with a large transverse energy ($E_{\mathrm{T}} > 25$~GeV) and $|\eta| < 2.5$, excluding the barrel--endcap transition region $1.37 < |\eta| < 1.52$, or one muon with large transverse momentum ($\pT > 20$~GeV) and $|\eta| < 2.5$ is required. The selected lepton
must match a lepton trigger object that caused the event to be recorded.
Jets present in $W/Z$+jet events tend to originate from soft gluon emissions.
These backgrounds are therefore reduced by requiring at least four jets with $\pT > 25$ GeV and $|\eta| < 2.5$.
At least two jets must be identified as originating from a $b$-decay 
using the MV1 algorithm.
To suppress backgrounds from QCD multi-jet events, the 
missing transverse momentum is required to be $\met > 20 (30)$~GeV in the muon (electron) channel.
Further reduction of the multi-jet background is achieved by requiring
the transverse mass\footnote{$m_{\mathrm{T}}=\sqrt{2p_{\mathrm{T}}^\ell\met(1-\cos\Delta\phi)}$ 
where $\Delta\phi$ is the azimuthal angle between the lepton 
and the missing transverse momentum} ($m_{\mathrm{T}}$) of the lepton and $\met$
to satisfy $m_{\mathrm{T}} > 30$~GeV 
in the electron channel and
$(\met + m_{\mathrm{T}}) > 60$~GeV in the muon channel. These requirements favour the presence of a $W$ boson, decaying
to $\ell \nu$, in the final state.
The selections are more stringent in the electron channel because of the larger
multi-jet background.

\section{Kinematic fit}
In the selected events, the two jets originating from the decay of 
the $H^+$ must be identified in order to reconstruct the mass. A kinematic fitter~\cite{cdfRef} is used to identify
and reconstruct the mass of dijets from $W/H^+$ candidates,
by fully reconstructing the $\ttbar$~system.
In the kinematic fitter, the lepton, $\met$ (assumed to be from the neutrino), 
and four jets are assigned to the decay particles from the $\ttbar$~system.
The longitudinal component of the neutrino momentum is calculated from the constraint
that the invariant mass of the leptonic $W$~boson decay products must be the experimental value ($80.4$~GeV)~\cite{pdg}. 
This leads to two possible solutions
for this momentum. When complex solutions are returned, the real part of the solution is used in the fit.
The fitter also constrains the invariant mass of the two systems ($b\ell\nu,bjj$) to be within $\Gamma_{t}=1.5$~GeV of the top-quark mass 172.5~\gev, which is consistent with the measured top-quark mass~\cite{Aaltonen:2012ra}.
When assigning jets in the fitter,
$b$-tagged jets are assumed to originate from the $b$-quarks.
The best $bbjj$ combination is found by minimizing a $\chi^2$ for each assignment of jets to quarks and for the choice of solution for the longitudinal neutrino momentum, where the five highest-$\pt$ jets are considered as possible top-quark decay products.
Since the $b$-jets are only allowed to be assigned to the $b$-quarks, and the two untagged jets are assigned to quarks from the same charged boson, there are two possible jet configurations overall for events with four jets, two of which are $b$-tagged.
For events with at least five jets, the two highest-$\pt$ jets are always assumed to be from the top-quark decay products ($W/H^+$ boson or $b$-quark) to reduce the combinatorics in the fit procedure.
The combination with the smallest $\chi^2$~value, $\chi^2_{\mathrm{min}}$, is selected as the best assignment.
The function minimized in the fit is:

\begin{equation}
\begin{aligned}
\label{cs_eq_chi2}
\chi^2 &= \sum_{i = \mathrm{\ell}, 4 \mathrm{jets}} \frac{(\pt^{i,\mathrm{fit}} - \pt^{i,\mathrm{meas}})^2}{\sigma_i^2} \\
        &+  \sum_{j = x,y} \frac{(p_j^{\mathrm{SEJ,fit}} - p_j^{\mathrm{SEJ,meas}})^2}{\sigma_{\mathrm{SEJ}}^2}\\
        &+ \sum_{k=jjb,b\ell\nu}\frac{ (m_k - m_{\mathrm{t}})^2}{\Gamma_{t}^2}~~.
\end{aligned}
\end{equation}

In the first term, the fitted transverse momenta of the lepton and the four jets currently under consideration are allowed to vary around the measured values using the corresponding measured resolutions ($\sigma_i$). In the fit only the magnitudes of the object $\pt$s are varied; the angles of the jets and leptons are assumed to be measured with good precision. The vector sum of the momenta of the remaining jets ($\pt>15$~GeV) in the event, labelled SEJ, is allowed to vary in the second term. The resolution for this term is taken from the nominal jet resolution.
Letting the SEJ vary allows the $\met$ to be recalculated from the fitted values of its dominant components. 
Jets with lower $\pt$ and energy from calorimeter cells not associated with any physics object are both minor 
contributions to the $\met$ and are held fixed in the re-calculation of the $\met$.
The third term constrains the hadronic ($jjb$) and leptonic ($b\ell\nu$) top-quark candidates to have a mass
close to the top-quark mass.

\begin{figure}[thbp]
\begin{center}
\includegraphics[width=0.99\columnwidth]{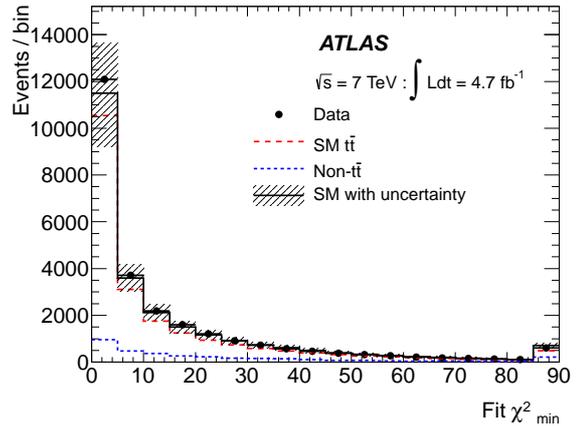}
\caption{\label{f:chi2}
Comparison of the distribution of $\chi^2_{\mathrm{min}}$ from the kinematic fitter
for data and the expectation from the background estimates
for the combined electron and muon channels.
The MC simulation is normalized to the expectation for the SM ($\Br(t\to H^+b)=0$).
The uncertainty shown on the background estimate is the combination in quadrature of the $\pm 1 \sigma$~systematic uncertainties.
The final bin also contains the overflow entries.}
\end{center}
\end{figure}

\begin{figure}[htbp]
\begin{center}
\includegraphics[width=0.99\columnwidth]{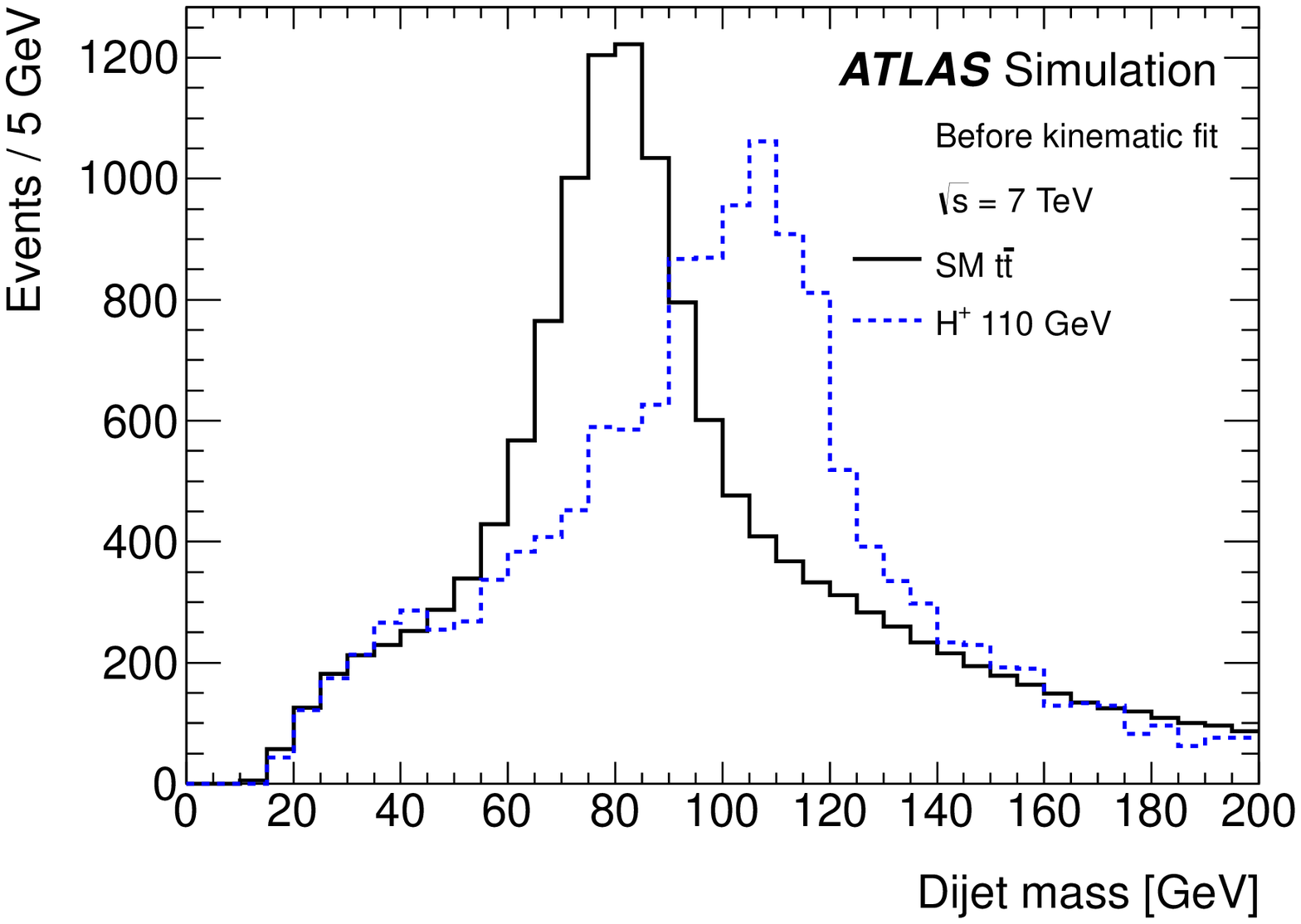}
\includegraphics[width=0.99\columnwidth]{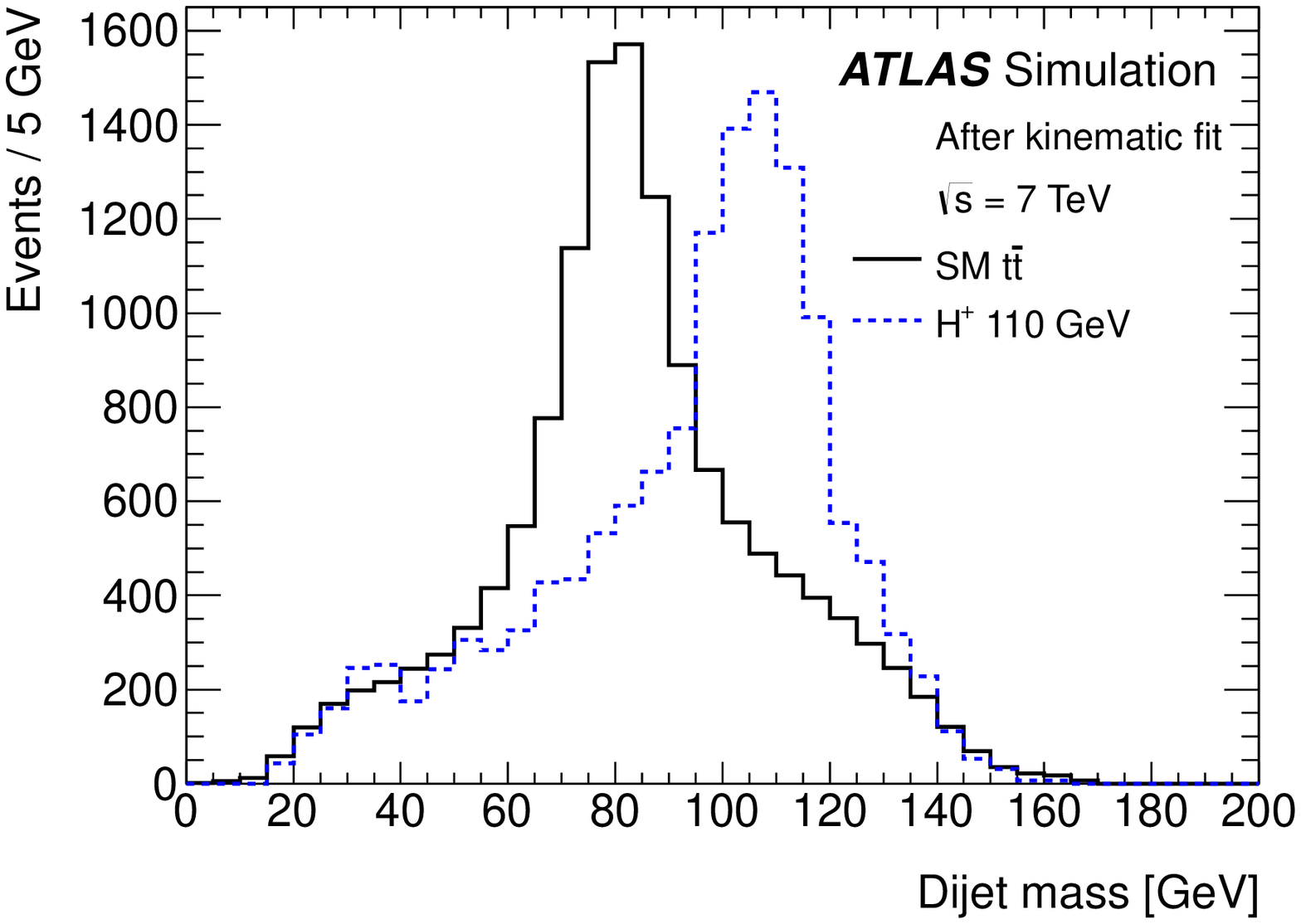}
\caption{
Comparison of the dijet mass distribution 
before (upper part) and after (lower part) the kinematic fit and the $\chi^2 < 10$ selection criterion. 
The distribution is shown for MC simulations
of SM $\ttbar$~decays and the $m_{H^+} = 110$~GeV signal ($\ttbar\rightarrow H^+bW^-\bar{b}$). The curves are normalized to the same area.
}
\label{cs_fit_templates}
\end{center}
\end{figure}

The $\chi^2_{\mathrm{min}}$~distribution for selected events in the data agrees well with the expectation from the simulation (see Fig.~\ref{f:chi2}).
Events are required to have $\chi^2_{\mathrm{min}} < 10$~to remove poorly reconstructed 
$\ttbar$~events.
This selection has an efficiency of $63\%$~for SM $\ttbar$~events.
The fit results in a 12 GeV dijet mass resolution,
 as shown in Fig.~\ref{cs_fit_templates}. This is a 20--30$\%$ improvement, 
depending on the mass of the boson studied, compared
to the resolution obtained when the same jets are used with their original
transverse momentum measurements. 
After the fit, there is better discrimination 
between the mass peaks of the $W$~boson 
from SM decays of $\ttbar$ and a $110$~GeV $H^+$~boson in this example.


Table~\ref{tab:evts} shows the number of events observed in the data and the number of events expected from the SM processes after the selection requirements. The SM $\ttbar$ entry includes events from both the lepton + jets and dilepton $\ttbar$~decay modes, where the dilepton events can pass the event selection if the events contain additional jets and the second lepton is not identified. Good agreement is observed between the data and the expectation. The table also shows the number of signal events expected for $\Br(t \rightarrow H^{+}b)=10\%$. The signal prediction accounts for acceptance differences due to the different kinematics of the $t\rightarrow H^+b$ events relative to the SM $t\rightarrow Wb$ events.

\begin{table}
\newcommand\T{\rule{0pt}{2.6ex}}
\newcommand\B{\rule[-1.2ex]{0pt}{0pt}}
\begin{center}
\begin{tabular}{ l  |  r@{\hspace{0.05cm}}l  | r@{\hspace{0.05cm}} l  }
\hline
Channel           & \multicolumn{2}{ c  |}{Muon} &  \multicolumn{2} 
{ c }{Electron} \\
\hline
Data              &  \multicolumn{2}{ c |}{$10107$}       &  
\multicolumn{2}{ c }{$5696$} \\
\hline
\T SM $\ttbar \rightarrow W^+b W^-\bar{b}$          & $8700$ &  
$\pm{1800}$ & $5000$ & ${\pm1000}$\\
$W/Z+{\rm jets}$    & $420$ & $\pm120$ & $180$ & $\pm50$ \\
Single top quark + Diboson  & $370$ & $\pm60$ & $210$ & $\pm30$ \\
QCD multi-jet      & $300$ & $\pm150$ & $130$ & $\pm60$   \\
\hline
\T \B Total Expected (SM)& $9800$ & $\pm{1800}$ & $5500$ &  $\pm{1000}$ \\
\hline\hline
$m_{H^+} = 110$~GeV &&\\
\hline
$\Br(t\to H^+b)=10\% :$&&\\
~~~$\ttbar \rightarrow H^+bW^-\bar{b}$  & $1400$ &  
$\pm{280}$     & $800$ & $\pm{160}$\\
\B ~~~$\ttbar \rightarrow W^+b W^-\bar{b}$ & $7000$ &  
$\pm{1400}$     & $4000$ & $\pm{800}$\\
\hline
\T \B Total Expected $(\Br=10\%)$ & $9500$ & $\pm{1700}$   & $5300$ & $\pm{1000}$\\
\hline
\end{tabular}
\caption{
The expected numbers of events from SM processes, integrated over the full range of dijet masses and
 the observed number of events in the data after all the selection requirements.
The expected number of events in the case of a signal with $m_{H^ 
+}=110$~GeV
and $\Br(t\to H^+b)=10\%$~is also shown. The $\ttbar\rightarrow W^+b W^-\bar{b}$ numbers include both the lepton + jets and dilepton decay channels. 
The uncertainties are the sum of the contributions from statistics and
systematic uncertainties.
\label{tab:evts}
}
\end{center}
\end{table}

\section{Systematic uncertainties}

The background estimates and the estimate of the signal efficiency are subject to a number of systematic uncertainties.
The QCD multi-jet background is estimated using a data-driven 
method~\cite{jet-electron} that employs a likelihood fit 
to the $\met$ distribution in the data, using a template 
for the multi-jet background
and templates from MC simulations for all other processes.
The uncertainty on the QCD multi-jet background 
is evaluated to be 50\% by studying the effect of pile-up events 
on the fit results and by performing likelihood fits on the $m_{\mathrm{T}}(W)$ distribution.
The dijet mass distribution of multi-jet events is obtained from a control region in the data, where leptons are required to be semi-isolated, such that the transverse momentum
of the inner detector tracks in a cone of radius $\Delta R=0.3$, excluding the lepton, satisfies $0.1<\pt^{\Delta R = 0.3}/\pt(e,\mu)<0.3$.
Leptons in the control region are also required to have a large impact
parameter with respect to the identified primary vertex (0.2~mm $<|d_0|<$ 2~mm) and an impact parameter significance 
$|d_0|/\sigma_{d_0}>3$.

The rate of $W$+jets events is estimated by a data-driven method~\cite{Wasymm} that uses the observed difference in the number of $W^+$ and $W^-$ bosons in the data and the charge asymmetry $(W^+-W^-)/(W^++W^-)$, which is calculated to good precision by the MC simulation of $W$+jets events. The heavy flavour fraction of the $W+$jets MC simulation is calibrated using $W+1$ jet or $W+2$ jets events in the data.
The uncertainty on the $W$+jets background is 26\% (28\%)
for the electron (muon) channel, which includes the uncertainty from the charge asymmetry and heavy flavour fraction components.
The shape of the $m_{jj}$ distribution for $W+$jets events 
is obtained from simulation.

Uncertainties on the modelling of the detector and on theory give rise to systematic uncertainties on the signal and background rate estimates.
The following systematic
uncertainties are considered: integrated luminosity ($3.9\%$)~\cite{Aad:2011dr, ATLAS-CONF-2011-116},
trigger efficiency ($3.5\%/1\%$ for electron/muon), 
jet energy scale (1--4.6\%)~\cite{Aad:2011he}, 
jet energy resolution (up to $16\%$ smearing)~\cite{Aad:2012ag}, 
and $b$-jet identification efficiency (5--17\%).
The last three uncertainties depend on the $\pT$~and $\eta$~of the jets.
Uncertainties on lepton reconstruction and identification efficiency are determined using a tag and probe method in samples of $Z$ boson and $J/\psi$ decays~\cite{MuonRecon}. The momentum resolution and scales are determined from fits to samples of $W$ boson, $Z$ boson, and $J/\psi$ decays~\cite{MuonResol,Aad:2011mk}.
Additional $\pt$-dependent uncertainties are placed on the $b$-jet 
(up to $2.5\%$) and $c$-jet (up to $1.3\%$) energy scales~\cite{Aad:2011he}.
Uncertainties on the modelling of the $\ttbar$~background are estimated using a second
MC generator ({\sc åÊPowheg}~\cite{Nason:2004rx, Frixione:2007vw, Alioli:2010xd}) and comparing the effect of using {\sc åÊPythia} and {\sc åÊHerwig} to
perform the parton showering and hadronization.
Uncertainties on initial and final state radiation (ISR/FSR) are assessed using {\sc åÊAcerMC} interfaced to
{\sc åÊPythia} and examining the effects of changing the ISR/FSR parameters in a range consistent with
experimental data~\cite{TopVeto}.
The predicted SM $\ttbar$ cross-section for $pp$ collisions at $\sqrt{s} = 7 \tev$, obtained from approximate next to next to LO QCD calculations, is $\sigma_{t\bar{t}} = 167^{+17}_{-18}$~pb for a top-quark mass of $172.5 \gev$~\cite{Aliev:2010}.
The uncertainty on the predicted value includes the uncertainty 
in the renormalization and factorization scales,
parton density functions, and the strong coupling constant.
An additional uncertainty on the $\ttbar$ cross-section ($4.5\%$) is included due to the uncertainty on the top-quark mass.
The uncertainty on the top-quark mass is 0.9 GeV from the combined measurement~\cite{Aaltonen:2012ra}  at the Tevatron.
However, this result would be biased in the presence of a $H^+\rightarrow\csbar$ signal in the lepton + jets channel, so a larger uncertainty of 1.5~GeV is taken, which is consistent with the latest top-quark mass measurement in the dilepton channel from the CMS experiment~\cite{CMSDilepton}. Changing the top-quark mass leads to altered event kinematics, which results in a final uncertainty on the event rate of $1.9\%$.
The effects of these systematic uncertainties on the overall normalization 
ÃÃare listed in Table~\ref{t:syst}.
The jet energy calibration, $b$-jet identification, $\ttbar$~background modelling, 
and ISR/FSR uncertainties
also modify the shape of the dijet mass distribution and are therefore 
determined as a function of $m_{jj}$. The systematic uncertainties that affect the shape of the $m_{jj}$ distribution (top half of Table~\ref{t:syst}) are more important than the shape-independent uncertainties. The effects of the systematic uncertainties are comparable, within 10\%, between the SM and signal $\ttbar$ samples.
The combined uncertainty on the single top-quark and diboson backgrounds
is 15\%, which comes mostly from the uncertainties 
on the cross-section, jet energy scale, and $b$-tagging.
The total uncertainty on the overall normalization of 
the non-$\ttbar$ backgrounds is 30\%.

\begin{table}
\begin{center}
\begin{tabular}{   c | c  }
\hline
Systematic Source & \\
\hline
\multicolumn{2}{c}{Shape dependent}\\
\hline
Jet energy scale                    & $\pm9.5 \%$  \\
$b$-jet energy scale                    & $+0.3, -0.6\%$\\
$c$-jet energy scale                    & $+0.1, -0.3\%$\\
Jet energy resolution                  & $\pm 0.9 \%$  \\
MC generator                           & $\pm 4.3 \%$ \\
Parton shower                          & $\pm 3.1 \%$ \\
ISR/FSR                                & $\pm 8.8 \%$\\
\hline
\multicolumn{2}{c}{Shape independent}\\
\hline
$b$-tagging efficiency ($b$-jets)                 & $\pm11\%$ \\
$b$-tagging efficiency ($c$-jets)                 & $\pm 2.4\%$ \\
$b$ mistag rate                 & $\pm 1.8\%$ \\
Lepton identification                 & $\pm 1.4 \%$  \\
Lepton reconstruction                & $\pm 1.0 \%$  \\
$t$-quark mass                & $\pm1.9\%$ \\
$t\bar{t}$~cross-section               & $+10, -11\%$ \\
Luminosity                             & $\pm 3.9 \%$   \\
\hline
\end{tabular}
\caption{\label{t:syst}
Effect of the systematic uncertainties on the event rate of $\ttbar$~background
and signal ($m_{H^+}=110$~GeV) events before any reduction from the likelihood fit, described in Sect.~\ref{results}.
}
\end{center}
\end{table}

\begin{figure}[]
\begin{center}
\includegraphics[width=\columnwidth]{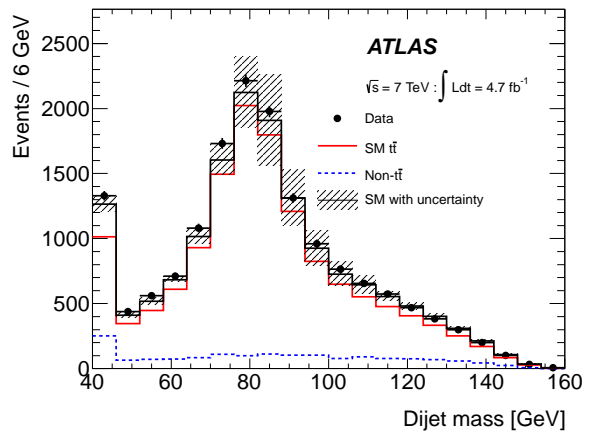}
\caption{\label{dijet_data}
The dijet mass distribution from data and the expectation from the SM ($\Br=0$).
The error bars represent the statistical uncertainty on the data. The uncertainty shown on the background estimate is the combination in quadrature of the $\pm 1 \sigma$~systematic uncertainties, accounting for the constraint from the profile likelihood fit.
The first and last bins contain the underflow and overflow events respectively.
}
\end{center}
\end{figure}

\section{Results}
\label{results}
The data are found to be in good agreement with the distribution of the dijet mass expected from SM processes (see Fig.~\ref{dijet_data}). The fractional uncertainty on the signal-plus-background model is comparable to the background only model. Upper limits on the branching 
ratio $\Br(t\rightarrow H^+b)$ are extracted as a function of the charged Higgs boson mass. The upper limits are calculated
assuming the charged Higgs always decays to $\csbar$.
The following likelihood function is used to describe the expected number of events as a function of the branching ratio:

\begin{equation}
\mathcal{L}(\Br, {\mathbf{\alpha}}) = \prod_i \frac{\nu_{i}(\Br,\alpha)^{n_i} e^{-\nu_{i}(\Br,\alpha)}}{n_i!} \prod_j \frac{1}{\sqrt{2\pi}} e^{-\frac{\alpha_j^2}{2}},
\label{eqn:lh}
\end{equation}

where $n_i$~is the number of events observed in bin $i$ of the dijet mass distribution and $j$~labels the sources of systematic uncertainty.
The number of expected signal plus background events in each bin, 
$\nu_{i}(\Br,\alpha)$,
 is given by

\begin{equation}
\begin{aligned}
\nu_{i}(\Br,\alpha)  = 
	2\Br(1-\Br) \, \sigma_{t\bar{t}}\,\mathcal{L}\, A^{H^{+}} S_i^{H^{+}} \prod_{j\ne b} \rho^{H^{+}}_{ji}( \alpha_j) \\
      + (1 - \Br)^2 \, \sigma_{t\bar{t}} \,\mathcal{L}\, A^{W} S_i^{W} \prod_{j\ne b} \rho^{W}_{ji}( \alpha_j) + n^N_{i}\rho^N_{bi}(\alpha_b)
\end{aligned}
\label{eqn:num}
\end{equation}

where $n^N_i$~is the expected number of non-$t\bar{t}$~background events,
$\sigma_{t\bar{t}}$~is the cross-section for $t\bar{t}$ production, $\mathcal{L}$~is the integrated luminosity,
$\Br$~is the branching ratio of $t\rightarrow H^+b$, 
and $A^{H^{+}}$ and $A^{W}$ are the acceptances for signal ($\ttbar\to H^+b\ell\nu\bar{b}$)~and SM $\ttbar$ ($\ttbar\to jjb\ell\nu\bar{b}$ and $\ttbar\to\ell\bar{\nu} b\ell\nu\bar{b}$)~events respectively.
The decay mode $t\bar{t}\rightarrow H^+bH^-\bar{b}$~does not contribute to the expectation because this mode does not produce
a single isolated lepton and hence has a negligible efficiency to pass the selection requirements.
The $S_i^{H^+}$ ($S_i^{W}$) parameter describes the shape
of the $m_{jj}$ spectrum (normalized to one) for $H^+$ ($W$) boson production.
It gives the relative number of events in bin $i$
according to the normalized $m_{jj}$ distribution.
The $\alpha_j$ variables are nuisance parameters representing 
the systematic uncertainties,
which are constrained via the Gaussian terms in Eq.~\ref{eqn:lh}. 
The effect of the systematic uncertainties on the non-$t\bar{t}$ background can be obtained by calculating the effect of each source of uncertainty on each non-$t\bar{t}$ background component, and combining them in quadrature. ÊSince this sum is dominated by the uncertainties on the data-driven $W$+jets and multi-jet background estimates, the combined variation is treated as a single nuisance parameter ($\alpha_b$, $b \in j$) and is assumed to be uncorrelated from the other systematic uncertainties.
The $\rho_{ji}$ functions account for the effect of nuisance parameters
on the yields and are defined such that
$\rho_{ji}(\alpha_j=\pm 1\sigma)$~represents the $1\pm 1\sigma$~fractional change in the number of entries in bin $i$ of the dijet mass spectrum due to systematic uncertainty $j$.
The physics measurement involves a sufficiently large number of events that this likelihood can constrain the $\alpha_j$ parameters beyond the precision of the subsidiary measurements.
The effects of systematic uncertainties are applied coherently in signal and background distributions. The subsidiary measurements of the $\alpha_j$ parameters are taken to be uncorrelated. The fit uses 17 nuisance parameters in total. None of them are shifted by more than one sigma compared to
the original values obtained in subsidiary measurements. Maximal reduction of uncertainty is obtained for the jet energy scale parameter which is reduced by 50\%.

\begin{figure}[]
\begin{center}
\includegraphics[width=\columnwidth]{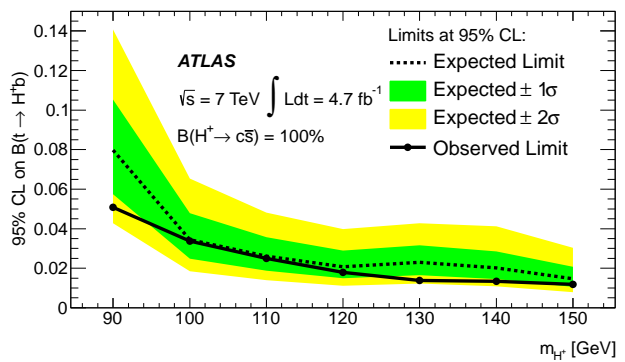}
\caption{\label{atlas_limits}
The extracted 95\% CL upper limits on ${\mathcal B}(t \rightarrow H^+b$), assuming that $\Br(H^{+}\rightarrow\csbar) = 100\%$, are shown for a range of charged Higgs masses from
90 GeV to 150 GeV. The limits shown are calculated using the CL$_{\mathrm{s}}$ limit-setting procedure.}
\end{center}
\end{figure}

The limits on the branching ratio are extracted using the CL$_{\mathrm{s}}$ technique at 95$\%$ confidence level~\cite{Cowan:2010js,CLs}.
The consistency of the data with the background model can be determined by comparing the
value of the test statistic (a profile likelihood ratio based on Eq.~\ref{eqn:lh}) in the data with the expectation from background-only Monte Carlo simulated experiments.
The corresponding probability ($p$-value) for the background to produce the observed mass distribution varies from $67\%$~to $71\%$ as a function of $m_{H^+}$, indicating that there is no significant deviation from the background hypothesis. 
The expected and observed limits, shown in Table~\ref{t:freqlimits} and Fig.~\ref{atlas_limits}, are calculated using asymptotic formulae~\cite{Cowan:2010js}.
The expected limits on $\mathcal{B}$, including both statistical and systematic uncertainties, vary between 1--8$\%$ depending
on $m_{H^+}$; if only the statistical uncertainty is considered these limits are 1--3$\%$. The observed limits, including both statistical and systematic uncertainties, vary between 1--5$\%$.
The extracted limits are the most stringent to date on the branching ratio $\Br(t\to H^+b)$, 
assuming $\Br(H^+\to c\bar{s})=100\%$. These results can be used to set limits for a generic scalar charged boson decaying to dijets in top-quark decays, as long as the width of the resonance formed is less than the experimental dijet resolution of 12~GeV.

\begin{table}
\begin{center}
\begin{tabular}{c | c |c}
\hline
Higgs Mass & Expected limit & Observed limit \\
   &  (stat.$\oplus$ syst.)   &   (stat.$\oplus$ syst.)   \\
\hline
90 GeV   & 0.080  & 0.051  \\
100 GeV   & 0.034  & 0.034   \\
110 GeV   & 0.026  & 0.025   \\
120 GeV   & 0.021  & 0.018  \\
130 GeV   & 0.023  & 0.014   \\
140 GeV   & 0.020  & 0.013   \\
150 GeV   & 0.015  & 0.012   \\
\hline
\end{tabular}
\caption{Expected and observed $95\%$~CL limits, including systematic uncertainties,
on the branching ratio for a top-quark to decay to a charged Higgs boson
and a $b$-quark, assuming that $\Br(H^{+}\rightarrow\csbar) = 100\%$. The limits shown are calculated using the CL$_{\mathrm{s}}$ limit-setting procedure.}
\label{t:freqlimits}
\end{center}
\end{table}

\section{Conclusions}

A search for charged Higgs bosons decaying to $\csbar$ in $\ttbar$ production has been presented.
The dijet mass distribution is in good agreement with the expectation from the SM
and limits are set on the branching ratio $\Br(t\to H^+b)$, 
assuming $\Br(H^+\to c\bar{s})=100\%$.
The observed limits range from
$\Br=5\%$~to $1\%$~for $m_{H^+}=90$~GeV~to~$150$~GeV. 
These are the best limits to date on charged Higgs boson production in this channel.


\section{Acknowledgements}

We thank CERN for the very successful operation of the LHC, as well as the
support staff from our institutions without whom ATLAS could not be
operated efficiently.

We acknowledge the support of ANPCyT, Argentina; YerPhI, Armenia; ARC,
Australia; BMWF and FWF, Austria; ANAS, Azerbaijan; SSTC, Belarus; CNPq and FAPESP,
Brazil; NSERC, NRC and CFI, Canada; CERN; CONICYT, Chile; CAS, MOST and NSFC,
China; COLCIENCIAS, Colombia; MSMT CR, MPO CR and VSC CR, Czech Republic;
DNRF, DNSRC and Lundbeck Foundation, Denmark; EPLANET, ERC and NSRF, European Union;
IN2P3-CNRS, CEA-DSM/IRFU, France; GNSF, Georgia; BMBF, DFG, HGF, MPG and AvH
Foundation, Germany; GSRT and NSRF, Greece; ISF, MINERVA, GIF, DIP and Benoziyo Center,
Israel; INFN, Italy; MEXT and JSPS, Japan; CNRST, Morocco; FOM and NWO,
Netherlands; BRF and RCN, Norway; MNiSW, Poland; GRICES and FCT, Portugal; MERYS
(MECTS), Romania; MES of Russia and ROSATOM, Russian Federation; JINR; MSTD,
Serbia; MSSR, Slovakia; ARRS and MVZT, Slovenia; DST/NRF, South Africa;
MICINN, Spain; SRC and Wallenberg Foundation, Sweden; SER, SNSF and Cantons of
Bern and Geneva, Switzerland; NSC, Taiwan; TAEK, Turkey; STFC, the Royal
Society and Leverhulme Trust, United Kingdom; DOE and NSF, United States of
America.

The crucial computing support from all WLCG partners is acknowledged
gratefully, in particular from CERN and the ATLAS Tier-1 facilities at
TRIUMF (Canada), NDGF (Denmark, Norway, Sweden), CC-IN2P3 (France),
KIT/GridKA (Germany), INFN-CNAF (Italy), NL-T1 (Netherlands), PIC (Spain),
ASGC (Taiwan), RAL (UK) and BNL (USA) and in the Tier-2 facilities
worldwide.

\bibliographystyle{spphys}       
\bibliography{EPJC_csbar_2012}
\clearpage
\onecolumn
\begin{flushleft}
{\Large The ATLAS Collaboration}

\bigskip

G.~Aad$^{\rm 48}$,
T.~Abajyan$^{\rm 21}$,
B.~Abbott$^{\rm 111}$,
J.~Abdallah$^{\rm 12}$,
S.~Abdel~Khalek$^{\rm 115}$,
A.A.~Abdelalim$^{\rm 49}$,
O.~Abdinov$^{\rm 11}$,
R.~Aben$^{\rm 105}$,
B.~Abi$^{\rm 112}$,
M.~Abolins$^{\rm 88}$
% ...
\bigskip
\\
% ...
$^{11}$ Institute of Physics, Azerbaijan Academy of Sciences, Baku, Azerbaijan\\
$^{12}$ Institut de F{\'\i}sica d'Altes Energies and Departament de F{\'\i}sica de la Universitat Aut{\`o}noma de Barcelona and ICREA, Barcelona, Spain\\
% ...
$^{21}$ Physikalisches Institut, University of Bonn, Bonn, Germany\\
% ...
$^{48}$ Fakult{\"a}t f{\"u}r Mathematik und Physik, Albert-Ludwigs-Universit{\"a}t, Freiburg, Germany\\
$^{49}$ Section de Physique, Universit{\'e} de Gen{\`e}ve, Geneva, Switzerland\\
% ...
$^{88}$ Department of Physics and Astronomy, Michigan State University, East Lansing MI, United States of America\\
% ...
$^{105}$ Nikhef National Institute for Subatomic Physics and University of Amsterdam, Amsterdam, Netherlands\\
% ...
$^{111}$ Homer L. Dodge Department of Physics and Astronomy, University of Oklahoma, Norman OK, United States of America\\
$^{112}$ Department of Physics, Oklahoma State University, Stillwater OK, United States of America\\
% ...
$^{115}$ LAL, Universit{\'e} Paris-Sud and CNRS/IN2P3, Orsay, France\\
% ...
\end{flushleft}

 
%

\end{document}